\input harvmac.tex
\vskip 1.5in
\Title{\vbox{\baselineskip12pt
\hbox to \hsize{\hfill}
\hbox to \hsize{\hfill}}}
{\vbox{
	\centerline{\hbox{$\alpha$-Symmetries, Coloured Dimensions
		}}\vskip 5pt
        \centerline{\hbox{and Gauge-String Correspondence
		}} } }

\centerline{Dimitri Polyakov\footnote{$^\dagger$}
{dp02@aub.edu.lb}}
\medskip
\centerline{\it Center for Advanced Mathematical Sciences}
\centerline{\it and  Department of Physics }
\centerline{\it  American University of Beirut}
\centerline{\it Beirut, Lebanon}
\vskip .3in
\centerline {\bf Abstract}
We propose a scenario of gauge-string correspondence
by relating  the SU(3) colour group
to hidden space-time isometries originating
from extra dimensions. These isometries ( $\alpha$-symmetries) 
are the special symmetries of
 RNS superstring theories
under global
non-linear space-time transformations. 
The vertex operators for the octet of gluons are constructed
by the procedure of ``photon painting'', that is,
with the SU(3) subgroup of the $\alpha$-symmetry generators
acting on a regular open string photon, so the corresponding
open string excitations are
in the adjoint of SU(3).
Remarkably, the operator algebra of these massless gluon vertices is
closed and possesses the full zigzag symmetry, crucial for the isomorphism
between open strings and QCD.
As a result, the scattering amplitudes of the constructed 
open string vertex operators have a field-theoretic rather than a stringy
structure, including the absense of standard tower of
massive intermediate states.Our model also suggests that the total number
of underlying hidden dimensions is three, with each extra dimension carrying
its appropriate SU(3) colour and anticolour.

 {\bf }
\Date{June 2008}
\vfill\eject
\lref\verl{H. Verlinde, Phys.Lett. B192:95(1987)}
\lref\bars{I. Bars, Phys. Rev. D59:045019(1999)}
\lref\barss{I. Bars, C. Deliduman, D. Minic, Phys.Rev.D59:125004(1999)}
\lref\barsss{I. Bars, C. Deliduman, D. Minic, Phys.Lett.B457:275-284(1999)}
\lref\lian{B. Lian, G. Zuckerman, Phys.Lett. B254 (1991) 417}
\lref\pol{I. Klebanov, A. M. Polyakov, Mod.Phys.Lett.A6:3273-3281}
\lref\wit{E. Witten, Nucl.Phys.B373:187-213  (1992)}
\lref\grig{M. Grigorescu, math-ph/0007033, Stud. Cercetari Fiz.36:3 (1984)}
\lref\witten{E. Witten,hep-th/0312171, Commun. Math. Phys.252:189  (2004)}
\lref\wb{N. Berkovits, E. Witten, hep-th/0406051, JHEP 0408:009 (2004)}
\lref\zam{A. Zamolodchikov and Al. Zamolodchikov,
Nucl.Phys.B477, 577 (1996)}
\lref\mars{J. Marsden, A. Weinstein, Physica 7D (1983) 305-323}
\lref\arnold{V. I. Arnold,''Geometrie Differentielle des Groupes de Lie'',
Ann. Inst. Fourier, Grenoble 16, 1 (1966),319-361}
\lref\self{D. Polyakov,  Int.J.Mod. Phys A20: 2603-2624 (2005)}
\lref\selff{D. Polyakov, Phys. Rev. D65: 084041 (2002)}
\lref\ampf{S.Gubser,I.Klebanov, A.M.Polyakov,
{ Phys.Lett.B428:105-114}}
\lref\malda{J.Maldacena, Adv.Theor.Math.Phys.2 (1998)
231-252, hep-th/9711200} 
\lref\sellf{D. Polyakov, Int. J. Mod. Phys A20:4001-4020 (2005)}
\lref\selfian{I.I. Kogan, D. Polyakov, Int.J.Mod.PhysA18:1827(2003)}
\lref\doug{M.Douglas et.al. , hep-th/0307195}
\lref\dorn{H. Dorn, H. J. Otto, Nucl. Phys. B429,375 (1994)}
\lref\prakash{J. S. Prakash, 
H. S. Sharatchandra, J.Math.Phys.37:6530-6569 (1996)}
\lref\dress{I. R. Klebanov, I. I. Kogan, A. M.Polyakov,
Phys. Rev. Lett.71:3243-3246 (1993)}
\lref\selfdisc{ D. Polyakov, hep-th/0602209, to appear
in IJMPA}
\lref\wittwist{E. Witten, Comm. Math.Phys.252:189-258 (2004)}
\lref\wittberk{ N. Berkovits, E. Witten, JHEP 0408:009 (2004)}
\lref\cachazo{ F. Cachazo, P. Svrcek, E. Witten, JHEP 0410:074 (2004)}
\lref\barstwist{ I. Bars, M. Picon, Phys.Rev.D73:064033 (2006)}
\lref\barstwistt{ I.Bars, Phys. Rev. D70:104022 (2004)}
\lref\selftwist{ D. Polyakov, Phys.Lett.B611:173 (2005)}
\lref\klebwitt{I. Klebanov, E.Witten, Nucl.Phys.B 556 (1999) 89}
\lref\klebgauge{C. Herzog, I. Klebanov, P. Ouyang, hep-th/0205100}
\lref\ampconf{A. M. Polyakov, Nucl.Phys.B486(1997) 23-33}
\lref\amplib{A. M. Polyakov, hep-th/0407209,in 't Hooft, G. (ed.):
50 years of Yang-Mills theory 311-329}
\lref\wit{E.Witten{ Adv.Theor.Math.Phys.2:253-291,1998}}
\lref\alfa{D.Polyakov, Int.J.Mod.Phys.A22:5301-5323(2007)}
\lref\ghost{D.Polyakov, Int.J.Mod.Phys.A22:2441(2007)}

\centerline{\bf Introduction}

The gauge-string correspondence is a profound
hypothesis and a promising approach to important long-standing
problems in QCD (such as quark confinement)
, relating the observables (physical vertex
operators) in string theory to local gauge invariant
operators in QCD. In  particular, such a correspondence identifies
open strings with thin tubes of gluon field lines, connecting hadrons,
so
the Wilson loop's expectation value $<W(C)>$ on the QCD side 
is identified with the partition function $Z(C)$
of the open string with the ends attached to the same contour $C$.
 Once such an isomorphism
holds, one could expect that correlation functions
of massless vertex operators in open string theory
are to reproduce QCD dynamics.
Such a string-theoretic framework would be particularly
efficient and natural to address the problem of confinement, as well as
other non-perturbative QCD dynamics.

In practice, however, things are far more complicated.
First of all, if an open string is to describe the gluon dynamics,
its spectrum has to contain 8 massless vector bosons.
It is well-known that a perturbative open string spectrum
has only one massless physical excitation, a photon, which has no colour.
Although this complication can be corrected ( somewhat artificially)
by introducing the appropriate Chan-Paton's factors 
(such as Gell-Mann matrices), this is only the beginning of sorrows.
The main and fundamental problem in identifying a QCD string, is that
a spectrum of a normal open string contains an infinite 
tower of massive states,
in addition to the photon. These states appear as intermediate poles
in any Veneziano amplitude (including the  scattering of 
massless gauge bosons), so there is no way to separate the
gluon dynamics from massive string modes. For this reason,
the 
Veneziano amplitude for the massless gauge bosons 
in superstring theory has little to do with the 
scattering amplitudes of gluons in QCD, since the latter
do not, of course, have anything like an infinite
set of indermediate massive states.
This complication has an underlying geometrical
reason.
That is, the standard open string theory lacks an important
symmetry 
 that, however, is present on the gauge theory side.
Namely, while the string theory is only invariant under reparametrizations
with positive Jacobians (that do not change the worldsheet orientation),
the Wilson loop is also invariant under the orientation change
 ~{\ampconf, \ampconf, \amplib}. 
Technically,  
the loop equation satisfied by $<W(C)>$, is the consequence 
of the zigzag symmetry. It is easy to check that  Ward identities
for perturbative string theory in flat space-time background
do {\bf not}  reproduce the loop equation, indicating the 
absense of the zigzag symmetry. 
In other words, a stringy description of QCD requires the 
presense of closed subalgebra of massless vertex operators
of gluons
and,  as a consequence,
 the field-theoretic behaviour of their scattering amplitudes
(including the absense of the intermediate poles in 
correlation functions and OPE).

In the supersymmetric case, this problem can be partially
solved by the AdS/CFT correspondence, which is a
special case of  the gauge-string correspondence 
( see e.g. ~{\malda, \wit, \ampf, \klebgauge})
However, the formalism developed in the AdS/CFT framework is incomplete.
First of all, it mainly works in the large $N$ limit, which
has little to do with realistic gauge group of QCD, such as $SU(3)$.
In addition, the AdS/CFT correspondence 
only gives an isomorphism between closed string vertex operators 
(such as a dilaton) and gauge-invariant observables on the QCD side,
such as $Tr(F^2)$. This correspondence is incomplete; for example, 
it can't be used to  calculate gluon scattering amplitudes in QCD since 
, in order to describe the emission of gluons in string theory,
one needs to use the ${\it open}$ string vertex operators
coupled to the  gauge potential $A_m$, rather
than to the gauge-invariant fields. 
As we noted above, there seems to be no 
 straightforward way to deduce such emission vertices
from standard open string theory, or from the AdS/CFT approach.
In this paper, we  address the problem of the gauge-string correspondence,
 by
constructing the  massless vertex operators
for the SU(3) octet of gluons and studying their scattering amplitudes.
The key element of our construction is the set of very peculiar
space-time symmetries (referred to as $\alpha$-symmetries)
present in non-critical RNS superstring theories ~{\alfa, \ghost}.
These symmetries are realised non-linearly and are closely
related to the presence of hidden space-time dimensions.
Typically, they mix the matter and the
ghost degrees of freedom in superstring theory; in particular,
the variation of the matter part of the worldsheet RNS action
is cancelled by that of the superconformal ghost part
under the $\alpha$-symmetry transformations .
The alpha-symmetry generators are the BRST-invariant picture-dependent
primary fields which can be classified in terms of the ghost 
cohomologies $H_n\sim{H_{-n-2}(n=1,2,...)}$ ~{\alfa}. Typically, for
each the  cohomology $H_n\sim{H_{-n-2}(n=1,2,...)}$ 
in $d$-dimensional non-critical string theory contains
$n+2$ generators, one $d$-vector and $n+1$ scalars.
For each given $n$ these operators induce the 
subset of $\alpha$ symmetries
stemming from separate hidden space-time dimension, enhancing the 
space-time symmetry group and increasing the effective space-time 
dimensionality
by one unit. The $\alpha$-symmetry generators from the first
$n$ cohomologies $H_k\sim{H_{-k-2}}(1\leq{k}\leq{n})$, combined
with the standard translation and rotation
generators
of  the SO(2,d-1) space-time isometry group
(for d space-time dimensions plus the Liouville direction)
extend the symmetry group to $SO(2,d-1+n)$.
In this paper, of special interest to us are the $\alpha$-symmetries
that do not mix with the standard space-time  Poincare generators, i.e.
subgroup the 
 $\alpha$-generators acting purely in extra dimensions.
 These generators correspond to isometries
of  the hidden dimensions
(other than those of the visible space-time)
Remrkably, it turns out that
 these generators, applied to the Hilbert space of physical
states of non-critical RNS superstring theory, form SU(3)
subgroup which is  translated into the gauge group of strong interaction
in the gauge-string correspondence. The vertex operators 
for the gluon emission then emerge in the adjoint representation
of the $SU(3)$ subgroup of the extra dimensional $\alpha$-generators,
building the ground for the isomorphism between $SU(3)$ QCD
and giving rise special isolated zigzag-invariant 
sector of open string theory which is described
in terms of nonzero ghost cohomologies
 and is detouched from  the standard Hilbert space 
 perturbative open string oscillations.
So in our approach, the QCD string emerges not as a separate string theory in  
a certain space-time background (such as AdS), but as a special isolated sector
of RNS superstring theory in flat background, related to isometries
of hidden dimensions. Interestingly, the total number of hidden
dimensions in our model turns out to be three, with each extra dimension
carrying a corresponding colour or anticolour quantum number.

This paper is organized as follows.
In the next section we review the classification of the $\alpha$-symmetries
in terms of the ghost cohomologies and their relation to the hidden space-time 
dimensions.
In the section 3 we concentrate on the 
extra dimensional subset of the $\alpha$-generators, not intertwining with 
the symmetry generators of visible space-time.
We show this subset of the $\alpha$-symmetries to form $SU(3)$ subgroup
and construct the octet of gluon vertex operators in the adjoint
of $SU(3)$ by acting with the appropriate $\alpha$-generators
on an open string photon.Next, we discuss the OPE structure of
 these vertex operators and the closedness of their operator algebra,
implying the underlying zigzag symmetry. In particular, the absense
of massive intermediate poles in their scattering amplitudes is explained
by the special property of the $\alpha$-symmetry generators
to annihilate the massive states (typically, if an $\alpha$-generator is 
applied to any  string excitation other than massless, 
a BRST trivial operator is produced).In other words,
the $\alpha$-symmetry works like an ``Occam's razor'',
shaving off the massive intermediate states and restoring 
the zigzag symmetry.
In the section 4 we calculate some of the $4$-point scattering amplitudes
of the gluon vertex operators and show them to have a field-theoretic 
structure, consistent with the zigzag symmetry (no massive poles).
In the concluding section we discuss the implications of
our results and the 
directions for the future work.

\centerline{\bf 2. $\alpha$-Symmetries, Ghost Cohomologies and Hidden
Dimensions}

In our recent works ~{\ghost, \alfa} we have shown that non-critical
RNS  superstring theories are invariant under the  set of unusual 
non-linear space-time transformations, not at all
evident from the structure of their worldsheet actions.
That is, consider the worldsheet 
action of $d$-dimensional RNS superstring theory
 given by
\eqn\grav{\eqalign{S={1\over{2\pi}}\int{d^2z}\lbrace
-{1\over2}{\partial{X_m}}\bar\partial{X^m}-{1\over2}
\psi_m\bar\partial\psi^m-{1\over2}\bar\psi_m\partial\bar\psi^m\rbrace
+S_{ghost}+S_{Liouville}\cr
S_{ghost}={1\over{2\pi}}\int{d^2z}\lbrace
b\bar\partial{c}+{\bar{b}}\partial{\bar{c}}+
\beta\bar\partial\gamma+\bar\beta\partial\bar\gamma\rbrace\cr
S_{Liouville}={1\over{4\pi}}\int{d^2z}
{\lbrace}\partial\varphi\bar\partial\varphi+\lambda\bar\partial\lambda+
{\bar\lambda}\partial\bar\lambda-F^2+2\mu_0be^{b\varphi}
(ib\lambda\bar\lambda-F)\rbrace
}}
where $\varphi,\lambda,F$ sre the components of the
 Liouville superfield, , $X^m,m=0,...,d-1$
are the space-time coordinates,$\psi^m,\bar\psi^m$ are
their worldsheet superpartners; $b,c,\beta,\gamma$
are the fermionic and bosonic (super)reparametrization ghosts bosonized as
\eqn\grav{\eqalign{b=e^{-\sigma};c=e^\sigma,\cr
\beta=e^{\chi-\phi}\partial\chi\equiv\partial\xi{e^{-\phi}};
\gamma=e^{\phi-\chi}\cr
}}

This action is obviously invariant under two global
$d$-dimensional space time symmetries -
Lorenz rotations and translations. 
 One can straightforwardly  check, however, that in addition to these 
obvious symmetries
the action (1) is also invariant under the following non-linear
global transformations, mixing the matter and the ghost sectors of the
theory ~{\alfa}:

\eqn\grav{\eqalign{\delta{X^m}=\epsilon
\lbrace\partial(e^\phi\psi^m)+2e^\phi\partial\psi^m\rbrace\cr
\delta\psi^m=\epsilon\lbrace{-}e^\phi\partial^2{X^m}
-2\partial(e^\phi\partial{X^m})\rbrace\cr
\delta\gamma=\epsilon{e^{2\phi-\chi}}{\lbrace}\psi_m\partial^2{X^m}
-2\partial\psi_m\partial{X^m}{\rbrace}\cr
\delta\beta=\delta{b}=\delta{c}=0}}

with the
  generator of (3) given by

\eqn\lowen{T=\int{{dz}\over{2i\pi}}{e^\phi}(\partial^2{X_m}\psi^m
-2\partial{X_m}\partial\psi^m)}

The integrand of (4) is a primary field of dimension 1,
i.e. a physical generator.
While it is not manifestly BRST-invariant (it
doesn't commute with the supercurrent terms of $Q_{brst}$)
 its BRST invariance can be restored by 
adding the appropriate b-c ghost
dependent  terms according to the prescription described in ~{\ghost}
The special property of this generator is that 
it is annihilated by $\Gamma^{-1}$ and
 has no analogues
at higher pictures, such as $0,-1$ and $-2$
(but has versions at higher positive pictures $+2,+3,...$
which can be obtained by the standard picture-changing) 
The physical operators with such a property
are referred to as the  positive ghost cohomology $H_1$,
 ~{ \alfa, \ghost}.
For the sake of completeness, we recall the definition
and basic properties of the ghost cohomologies discussed in ~{\alfa, \ghost}.
The positive number $n$ ghost cohomology $H_n$ ($n=1,2,...$)
consists of physical (BRST invariant and non-trivial)
vertex operators violating the picture equivalence, 
 existing at picture $n$ and above, that are annihilated
at their minimal positive picture $n$ by the inverse picture changing 
operator $\Gamma^{-1}=c\partial\xi{e^{-2\phi}}$
(higher than $n$ pictures of such operators are related by 
the usual picture changing).
Similarly, the negative number $-n$ ghost cohomology $H_{-n}$ ($n\geq{3}$)
consists  of physical (BRST invariant and nontrivial) operators
that exist at picture $-n$ or below, that are 
annihilated by the direct picture changing
operator $\Gamma=:e^\phi{G}:$  at the $minimal$ negative picture $-n$
(here $G$ is the full matter $+$ ghost worldsheet supercurrent).
The operators of $H_{-n}$ at lower than $-n$ pictures
are related by the usual picture-changing.
There is an isomorphism between positive and negative ghost cohomologies
$H_{n}\sim{H_{-n-2}}$ as any element of $H_{-n-2}$ 
(typically having the form ${\sim}e^{-(n+2)\phi}F_{matter}$ at
 the minimal negative picture)
has a representation
in $H_n$
 obtained by replacing $e^{-(n+2)\phi}\rightarrow{e^{n\phi}}$
(with the matter part unchanged)
and adding the $b-c$ ghost counterterms (by the prescription 
described in ~{\ghost})
in order to protect the BRST invariance.
The usual picture-independent observables, existing at all pictures, including
picture $0$ (at which the superconformal ghosts decouple) are by definition
the elements of $H_0$. The cohomologies $H_{-1}$ and $H_{-2}$ are empty.

In particular, as the $\alpha$-symmetry generator (4) is the element of
$H_{1}$,
there is also a picture $-3$ version of this
generator, with the manifest BRST invariance. This version can be 
obtained simply by replacing $e^\phi\rightarrow{e^{-3\phi}}$ in (4).
Similarly to the picture $+1$-version, the picture ${-3}$
version is annihilated by $\Gamma$, so there are no versions of this
operator at pictures -2,-1 and 0 while the versions at pictures
below $-3$ can be obtained by straightforward inverse picture changing.

For this reason, the picture $-3$ version of (3) is an element
of negative ghost cohomology $H_{-3}$.

It is not difficult to show
 that, just like the operator (3) (the element of $H_{1}$)
generates the space-time symmetry transformations (2) of the RNS action (1),
similarly the picture $-3$ generates the 
symmetry transformations identical to (2)
(with $e^\phi$ replaced by $e^{-3\phi}$).
For the critical ($d=10$) uncompactified
 RNS superstrings the transformations (2) is the only
additional (to translations and rotations) space-time symmetry
generated by $H_1\sim{H_{-3}}$ currents.
For non-critical strings ($d\neq{10}$), however, there are
$d+1$ additional $\alpha$-symmetries, involving the Liouville sector.
(one $d$-vector and one scalar).

The corresponding generators are given by 

\eqn\grav{\eqalign{L^{m\alpha}=\oint{{dz}\over{2i\pi}}
e^{\phi}\lbrace\partial^2\varphi\psi^m
-2\partial\varphi\partial\psi^m+\partial^2{X^m}\lambda
-2\partial{X^m}
\partial\lambda\rbrace}}
and
\eqn\lowen{L^{\alpha-}=\oint{{dz}\over{2i\pi}}e^{\phi}
\lbrace\partial^2\varphi\lambda-
2\partial\varphi\partial\lambda\rbrace}
For simplicity, the expressions for $L^{m\alpha}$
and $L^{\alpha-}$ are given in the limit of zero 
cosmological constand and zero dilaton field so one can ignore the effect of 
the background charge; the corresponding expressions accounting for 
the Liouville dressing, are somewhat longer and  given in ~{\alfa}.

The appropriate space-time transformations are given by 
\eqn\grav{\eqalign{\delta{X_m}=\epsilon_{m\alpha}{\lbrace}
\partial(e^\phi\lambda)+2e^{\phi}\partial\lambda
\rbrace\cr
\delta\lambda
=-\epsilon_{m\alpha}{\lbrace}
2\partial(e^\phi\partial{X^m})+e^\phi
\partial^2{X^m}\rbrace\cr
\delta\gamma=\epsilon_{m\alpha}e^{2\phi-\chi}\lbrace
\partial^2{X^m}\lambda
-2\partial{X^m}\partial\lambda\rbrace\cr
\delta\beta=\delta{b}=\delta{c}=\delta\varphi=\delta\psi^m=0}}

and

\eqn\grav{\eqalign{\delta\varphi
=\epsilon_{-\alpha}\lbrace
\partial(e^\phi\lambda)+2e^\phi\partial\lambda
\rbrace\cr
\delta\lambda=-\epsilon_{-\alpha}\lbrace{2}\partial(e^\phi\partial\varphi)
+e^\phi{\partial^2}\varphi\rbrace\cr
\delta\gamma=\epsilon_{-\alpha}e^{2\phi-\chi}\lbrace
\lambda\partial^2\varphi-2\partial\varphi\partial\lambda
\rbrace\cr
\delta\beta=\delta{b}=\delta{c}=\delta{X^m}=\delta\psi^m=0}}.

The generators (5),(6) are the Virasoro primaries, annihilated
by the inverse picture changing (or by the direct p.c., if they
are taken in the $H_{-3}$ representation)
 They are BRST-invariant
(upon adding the $b-c$ ghost correction terms which we have skipped)
 and therefore
are the elements of $H_1\sim{H_{-3}}$. As before, the $H_{-3}$ version
of the generators (5),(6) with the manifest
BRST-invariance (with no $b-c$ correction terms)
can be obtained simply by replacing
$\phi\rightarrow{-3\phi}$ in (5) - (8).
Combined with ${{(d+1)(d+2)}\over2}$ dimension 1 Virasoro primaries:
\eqn\grav{\eqalign{L^{mn}=\oint{{dz}\over{2i\pi}}\psi^{m}{\psi^n}\cr
L^{+m}=\oint{{dz}\over{2i\pi}}e^{-\phi}\psi^m\cr
L^{-m}=\oint{{dz}\over{2i\pi}}\psi^m\lambda\cr
L^{+-}=\oint{{dz}\over{2i\pi}}e^{-\phi}\lambda}}

inducing the $d+1$ translations and ${{d(d+1)}\over2}$
rotations in space-time (including the Liouville direction),
the $d+2$ currents (4) - (6) of $H_1\sim{H_{-3}}$ 
 enlarge the current algebra of the space-time symmetry generators from
$SO(d,2)$ to $SO(d+1,2)$, effectively bringing an extra 
dimension to the theory.
Namely,
introducing the $d+2$-dimensional index
$M=(m,+,-,\alpha);m=0,...{d-2};\alpha=1$ with the $(d,2)$ metric
$\eta^{MN}$ consisting of $\eta^{mn},\eta^{+-}=-1,
\eta^{--}=\eta{++}=0,\eta^{\alpha\alpha}=1$
and evaluating the commutators of the operators (4) - (6)
and (9) it isn't difficult
to check that that
\eqn\lowen{\lbrack{L^{M_1N_1},L^{M_2N_2}\rbrack}
=\eta^{M_1M_2}L^{N_1N_2}+\eta^{N_1N_2}L^{M_1M_2}
-\eta^{M_1N_2}L^{M_2N_1}-\eta^{M_2N_1}L^{M_1N_2}}

Note that, as the $SO(d,2)$ group of translations and rotations for
 non-critical RNS strings is identical to the isometry
group of the $AdS_d$ space,  the $H_{1}\sim{H_{-3}}$ generators (4)-(6)
from the first nonzero ghost cohomology are simply the stringy
analogues of the off-shell symmetry generators from hidden
space-time  dimension, observed by Bars in the 2T physics approach
 ~{\bars, \barss, \barsss}.

It is straithforward to generalize the results to   the case
of the higher ghost cohomologies of the orders 2 and 3.
The general rule is that each ghost cohomology
$H_n\sim{H_{-n-2}};n=1,2,3$  contains the space-time  $\alpha$-symmetry
generators that induce the global 
space-time transformations leaving the RNS action invariant,
with the variation of the matter part cancelled by that of the ghost part.
Remarkably, each $H_n\sim{H_{-n-2}}$ cohomology pair contains
$d+n+1$ space-time symmetry generators, including one $d$-vector
and $(n+1)$ space-time scalars. Combining
the collection of the $\alpha$-generators of of $H_k\sim{H_{-k-2}};
1\leq{k}\leq{n}$ with the ordinary Poincare generators (including the Liouville
direction), the space-time isometry group is increased from
$SO(2,d)$ to $SO(2,d+n)$, i.e. $n$ hidden dimensions are induced, with
 each 
$H_k$ associated with an extra dimension.
The $H_2\sim{H_{-4}}$ $\alpha$-generators 
(those giving rise to the second extra dimension)
are given by:
\eqn\grav{\eqalign{
L^{\beta{+}}=\oint{{dz}\over{2i\pi}}e^{-4\phi}F_1(X,\psi)
F_1(\varphi,\lambda)(z)\cr
L^{\beta-}=-\oint{{dz}\over{2i\pi}}{e^{-4\phi}}
F_{1m}(X,\lambda)F_1^m(\varphi,\psi)(z)\cr
L^{\beta{m}}=\oint{{dz}\over{2i\pi}}e^{-4\phi}
(F_1^m(X,\lambda)F_1(\varphi,\lambda)
-F_1(X,\psi)F_1^m(\varphi,\psi))(z)\cr
L^{\alpha\beta}=\oint{{dz}\over{2i\pi}}
e^{-4\phi}({1\over2}F_2(\lambda,\varphi)+
L_1(X,\psi)\partial{L_1}(\varphi,\lambda)-\partial{L_1}(X,\psi)
L_1(\varphi,\lambda))(z)}}
with  the matter$+$Liouville structures $L$ and $F$
($L_1,F_1$ and $F_1^m$) being the primary fields of dimensions 2 and $5\over2$:
\eqn\grav{\eqalign{F_1(X,\psi)=
\psi_m\partial^2{X^m}-2\partial\psi_m\partial{X^m}\cr
F_1(\varphi,\lambda)=\lambda\partial^2\varphi-
2\partial\lambda\partial\varphi\cr
F_1^m(X,\lambda)=\lambda\partial^2{X^m}-2\partial\lambda\partial{X^m}\cr
F_1^m(\varphi,\psi)=\psi^m\partial^2\varphi-2\partial\psi^m\partial\varphi\cr
L_1(X,\psi)=\partial{X_m}\partial{X^m}-2\partial{\psi_m}\psi^m\cr
L_1(\varphi,\lambda)=(\partial\varphi)^2-2\partial\lambda\lambda}}
and $F_2(\lambda,\varphi)$ being the primary field of dimension 5:
\eqn\grav{\eqalign{
F_2(\varphi,\lambda)
={1\over4}(\partial\varphi)^5
-{3\over4}\partial\varphi(\partial^2\varphi)^2+{1\over4}
(\partial\varphi)^2\partial^3\varphi\cr+
\lambda\partial\lambda(\partial^3\varphi-(\partial\varphi)^3)
-{3\over2}\lambda\partial^2\lambda\partial^2\varphi
+3\partial\lambda\partial^2\lambda\partial\varphi
\rbrace\cr\equiv
i:(\oint{e^{-i\varphi}\lambda})^3{e^{3i\varphi}\lambda}:}}.
Combined with the matter $+$ Liouville Poincare generators of $SO(2,d)$
and the $\alpha$-generators (4) - (6) of $H_1\sim{H_{-3}}$,
the $\alpha$-generators of $H_2\sim{H_{-4}}$ satisfy the commutation relations
(10) of $SO(2,n+2)$ with the $(M,N)$ space-time indices supplemented by yet
another hidden dimension labelled by $\beta$:
$M= (m,\pm, \alpha,\beta)$.
Finally, the $\alpha$-generators at the level 
$H_3\sim{H_{-5}}$ (bringing in the third hidden dimension labelled by $\gamma$)
 are constructed as

\eqn\grav{\eqalign{L^{\gamma{+}}=\oint{{dz}\over{2i\pi}}
e^{-5\phi}{\lbrace}
2\partial{F_1}(X,\psi)-F_1(X,\psi)\partial{F_2}(\varphi,\lambda)\rbrace
\cr
L^{\gamma{m}}=\oint{{dz}\over{2i\pi}}e^{-5\phi}\lbrace
2F_2^m(\psi,\lambda,\varphi)\partial{F_1}(X,\psi)
-\partial{F_2}(\psi,\lambda,\varphi)F_1(X,\psi)\cr
+2F_2(\varphi,\lambda)\partial{F_1^m}(X,\lambda)
-\partial{F_2}(\varphi,\lambda)F_1^m(X,\lambda)\rbrace\cr
L^{\gamma-}=\oint{{dz}\over{2i\pi}}{e^{-5\phi}}
\lbrace
2G_2(\psi,\lambda,\varphi)\partial{F_1}(X,\psi)
-\partial{G_2}(\psi,\lambda,\varphi){F_1}(X,\psi)\cr
+3F_{2m}(\psi,\lambda,\varphi)\partial{F_1^m}(X,\lambda)
-2\partial{F_{2m}}(\psi,\lambda,\varphi)F_1^m(X,\lambda)
-\partial{F_2}(\lambda,\varphi)F_1(X,\psi)\rbrace
\cr
L^{\gamma\beta}=\oint{{dz}\over{2i\pi}}
e^{-5\phi}{\lbrace}F_3(\varphi,\lambda)
+\partial{L_1}(X,\psi)L_2(\varphi,\lambda)-{4\over{11}}L_1(X,\psi)
\partial{L_2}(\varphi,\lambda)\rbrace\cr
L^{\gamma\alpha}=\oint{{dz}\over{2i\pi}}e^{-5\phi}
{L_{2m}}(\varphi,\psi)L_1^m(X,\lambda)}}
with the additional matter$+$Liouville blocks given by:
\eqn\grav{\eqalign{F_2^m(\psi,\lambda,\varphi)=
\partial^2\psi^m\lambda\partial^2\varphi-\psi^m\partial^2\lambda
\partial^2\varphi
+3\partial^2\psi^m\partial\lambda\partial\varphi
-3\partial\psi^m\partial^2\lambda\partial\varphi\cr
G_2(\psi,\lambda,\varphi)=
4\partial\psi_m\partial^2\psi^m\partial\varphi-2\psi_m\partial^3\psi^m
\partial\varphi+(2d-4)(\lambda\partial^3\lambda\partial\varphi-
2\partial\lambda\partial^2\lambda\partial\varphi)\cr
L_2(\varphi,\lambda)=-{5\over4}(\partial\varphi)^4\partial\lambda
+{3\over4}(\partial^2\varphi)^2\partial\lambda
+{3\over2}\partial\varphi\partial^2\varphi\partial^2\lambda
-{5\over2}\partial\varphi\partial^3\varphi\partial\lambda\cr
-{1\over4}(\partial\varphi)^2\partial^3\lambda
-4\partial\varphi\partial^2\varphi\partial^2\lambda
+\partial^2\varphi\partial^3\varphi\lambda\cr
L_2^m(\varphi,\psi)=-{5\over4}(\partial\varphi)^4\partial\psi^m
+{3\over4}(\partial^2\varphi)^2\partial\psi^m
+{3\over2}\partial\varphi\partial^2\varphi\partial^2\psi^m
-{5\over2}\partial\varphi\partial^3\varphi\partial\psi^m\cr
-{1\over4}(\partial\varphi)^2\partial^3\psi^m
-4\partial\varphi\partial^2\varphi\partial^2\psi^m
+\partial^2\varphi\partial^3\varphi\psi^m\cr
L_1^m(X,\lambda)=\partial^2\lambda\psi^m+\lambda\partial^2\psi^m\cr
F_3(\varphi,\lambda)=:(\oint{e^{-i\varphi}\lambda})^4
{e^{-5\phi+4i\varphi}\lambda}:
}}
Combined with the space-time Poincare generators (9)
and with the $\alpha$-generators of two lower ghost cohomologies,
the $\alpha$-generators (14),(15) of $H_{-3}\sim{H_{-5}}$ extend
the space-time isometry group to $SO(2,d+3)$, unsealing the third hidden
dimension.
The problem of finding the BRST non-trivial $\alpha$-symmetry generators
in the higher order ghost cohomologies is more complicated,
as the expressions for the matter primaries of higher 
conformal dimensions become long and cumbersome.
The only symmetry generators
in the ghost cohomologies $H_{n}\sim{H_{-n-2}}$
for $n>3$ are those mixing the ghost and the Liouville sectors,
which can be obtained as the normal ordering of the operator
$\sim:(\oint{e^{-i\varphi}\lambda})^n{e^{-n\phi+i(n-1)\varphi}}\lambda:$
but such operators commute with the photon (thus producing
no new gauge bosons) and do not seem to have any 
straightforward analogues in the matter
sector.
At present, we do not know of any elegant  and compact 
prescription to construct such operators and, even if such higher order
$\alpha$-symmetries
exist and can be found, there is no evidence they could be interpreted 
in terms of extra dimensions, like those of the first three cohomologies.
On the contrary, if one is able to show that  
there are no
$\alpha$-symmetries (pointing at extra space-time dimensions)
in the ghost cohomologies $H_{n}\sim{H_{-n-2}}$ for $n>3$,
this may be an interesting
 interpretation of the $SU(3)$ strong interaction group
in the language of
extra dimensions. 
At present, at least, there is no compelling
pro or contra evidence for the $\alpha$-symmetries
(and related hidden space-time dimensions) originating from the 
ghost cohomologies at levels higher than 3.

\centerline{\bf 3. Construction of gluon emission vertices}

In this section, we 
 concentrate on the subgroup of the
the $\alpha$-symmetry generators (4) - (6),
(11) - (15) from the first three ghost
cohomologies $H_{n}\sim{H_{-n-2}}(n=1,2,3)$ 
 not mixing with 
Poincare transformations in visible space-time dimensions,
i.e. the generators without the space-time index $m$.
 We shall
 use this subgroup of generators to construct the gluon emission vertices.
There are altogether 9 generators:
$L^{\pm\alpha},L^{\pm\beta},{L^{\pm\gamma}}$
and $L^{\alpha\beta},L^{\alpha\gamma},L^{\beta\gamma}$.
 Physically, these generators correspond to the $\alpha$-isometries
of  hidden space-time dimensions.
Our goal now is to investigate how these $\alpha$-generators
act  on the massless open superstring state, i.e. the photon.
Generically, by acting on a photon with 9 scalar $\alpha$-generators
of the first three ghost cohomologies,
one would expect to construct a multiplet of 9 vertex operators
of massless vector
gauge bosons of $H_n\sim{H_{-n-2}} (n=1,2,3)$,
 related by the $\alpha$-transformations. 
Remarkably, however, one of the generators,
$L^{\alpha-}$, drops out as it turns out to commute with the photon
vertex (see the proof below). At the same time 
the remaining 8 generators do not commute with the photon,
producing 8 new vertex operators (massless gauge bosons), 
with each of them
inheriting the ghost cohomlogies of appropriate $\alpha$-transformations.
Below we shall construct explicitly 8 linear combinations
of the $\alpha$-generators generating $SU(3)$, so 
8 massless gauge bosons, obtained by $\alpha$-transforming the 
photon operator, will be in the adjoint of SU(3).
 
Since $L^{\alpha-}$ annihilates the photon, one can set
\eqn\lowen{L^{\alpha-}\approx{0}}
 as the subgroup of 9 scarar $\alpha$-generators
is applied to the RNS Hilbert space.
In particular this implies that any of the remaining 8 generators  
can be shifted by an operator proportional to $L^{\alpha-}$.
Given the constraint (16) it is now straightforward to show that
SU(3) is induced  by the following
eight linear combinations of the $\alpha$-generators of  
$H_{n}\sim
{H_{-n-2}};n=1,2,3$:

\eqn\grav{\eqalign{
F_{+}=-{1\over{{\sqrt{2}}}}(L^{\gamma+}+L^{\gamma-})
 -{i\over{{\sqrt{2}}}}(L^{\beta+}+L^{\beta-})+L^{\alpha\beta}-iL^{\alpha\gamma}
-{{i}\over{\sqrt{2}}}(L^{\alpha+}-L^{\alpha-})\cr
F_{-}=
-{1\over{{\sqrt{2}}}}(L^{\gamma+}+L^{\gamma-})
 -{i\over{{\sqrt{2}}}}(L^{\beta+}+L^{\beta-})-L^{\alpha\beta}+iL^{\alpha\gamma}
-{{i}\over{\sqrt{2}}}(L^{\alpha+}-L^{\alpha-})\cr
F_3=-{1\over{{\sqrt{2}}}}(L^{\gamma+}-L^{\gamma-})-{i\over{{\sqrt{2}}}}
(L^{\beta+}-L^{\beta-})\cr
L_1={i\over2}L^{\beta\gamma}\cr
L_2={i\over{{\sqrt{2}}}}(L^{\alpha+}+L^{\alpha-})\cr
G_{+}=-{1\over{{\sqrt{2}}}}(L^{\gamma+}+L^{\gamma-})
 +{i\over{{\sqrt{2}}}}(L^{\beta+}+L^{\beta-})+L^{\alpha\beta}+iL^{\alpha\gamma}
+{{i}\over{\sqrt{2}}}(L^{\alpha+}-L^{\alpha-})\cr
G_{-}=-{1\over{{\sqrt{2}}}}(L^{\gamma+}+L^{\gamma-})
 +{i\over{{\sqrt{2}}}}(L^{\beta+}+L^{\beta-})-L^{\alpha\beta}-iL^{\alpha\gamma}
+{{i}\over{\sqrt{2}}}(L^{\alpha+}-L^{\alpha-})\cr
G_3=-{1\over{{\sqrt{2}}}}(L^{\gamma+}-L^{\gamma-})+{i\over{{\sqrt{2}}}}
(L^{\beta+}-L^{\beta-})
}}
Here $L_1$ and $L_2$ are the Cartan generators,
three operators $F_{\pm}$ and $F_3$ are in the lowering 
subalgebra of $SU(3)$ while $G_{\pm}$ and $G_3$ are in the raising subalgebra.
The commutators of the operators (17) are straightforward to compute by using
the commutation relations (10) and imposing the constraint (16) 
upon the computation.
The next step is to analyse the transformations
of  the photon emission vertex  by this octet of generators.
 Since all the generators (17) are the integrated
space-time scalars of conformal dimension one, the result must involve
the octet of dimension 1 vertex operators - the gauge vector bosons
(provided there is no annihilation or BRST triviality).
We start with the  proof of the claim concerning
the annihilation of a photon by $L^{\alpha-}$.
The photon vertex operator taken at pictures $0$ and $-1$ is given by:

\eqn\grav{\eqalign{V_{ph}^{(0)}(k)=A_m(k)\oint{{dz}}
(\partial{X^m}+i({\vec{k}}{\vec{\psi}})\psi^m)e^{ik_nX^n}(z)\cr
V_{ph}^{(-1)}(k)=A_m(k)\oint{dz}e^{-\phi}\psi^m{e^{ik_nX^n}}(z)
;
\cr
m=0,1,2,3}}

with the $z$ integral taken over the worldsheet boundary.

If $L^{\alpha-}$ ( the element of $H_1\sim{H_{-3}}$)
 taken at picture $-3$ or below
 (i.e. in the $H_{-3}$ representation), 
, the annihilation
of $V_{ph}$ is obvious since the former depends only
on the Liouville and the ghost fields while $V_{ph}$ involves the matter.
Things, however, are not as straightforward if 
$L^{\alpha-}$ is taken at picture $+1$ or higher
(i.e. in the $H_1$ representation), 
since the OPE's involving the operators fron nonzero cohomologies are
generally picture-dependent ~{\ghost}. Still, it is easy
to check that the annihilation of
photon by $L^{\alpha-}$ holds for the $H_1$ representation as well.
The full BRST-invariant 
picture $+1$ representation of $L^{\alpha-}$ can be obtained
from its picture $-3$ version by changing $-3\phi\rightarrow\phi$
and adding the $b-c$ ghost dependent correction terms in order to protect
its BRST invariance, using the prescription described in ~{\ghost}, so
the complete expression for the picture $+1$ representation of
$L^{\alpha-}$ is

\eqn\grav{\eqalign{L^{\alpha-}_{+1}=
\int{{dz}\over{2i\pi}}e^\phi(\lambda\partial^2\varphi-
2\partial\lambda\partial\varphi)(z)\cr
+{1\over{8}}\oint_{w}{{dz}\over{2i\pi}}(z-w)^2
ce^\chi{\lbrace}(\psi_m\partial{X^m})(\lambda\partial^2\varphi-
2\partial\lambda\partial\varphi)(z)+U_{L}(z)\rbrace+...
\cr
\equiv
\int{{dz}\over{2i\pi}}e^\phi(\lambda\partial^2\varphi-
2\partial\lambda\partial\varphi)(z)
+{1\over{8}}\oint_{w}{{dz}\over{2i\pi}}(z-w)^2{\lbrace}(\psi_m\partial{X^m})
F_L(z)+ce^\chi{U_L}(z)\rbrace+...}}

where the second contour integral in (19)
 is taken around an arbitrary point $w$,
  $U_L$ is the dimension 4 operator
which depends only on the super liouville fields $\varphi$ and $\lambda$
(and therefore the corresponding term does not contribute to the commutator
of $L^{\alpha-}_{+1}$ with $V_{ph}^{(0)}$) and we denoted the 
ghost-Liouville dependent operator
$F_L(z)=c{e^\chi}(\lambda\partial^2\varphi-
2\partial\lambda\partial\varphi)(z)$ for the sake of brevity.
We also have skipped the correction terms proportional to 
$\sim(z-w)^2e^{\phi}(...)$ of $b-c$ ghost number zero,
as these terms depend on Liouville and ghost fields only,
automatically commuting with the picture zero photon.

The evaluation of the commutator of $L^{\alpha-}$ with the photon at 
picture zero gives:

\eqn\grav{\eqalign{{\lbrack}L^{\alpha-}_{+1},V_{ph}^{(0)}\rbrack\cr
=\oint{du}\oint_w{{dz}\over{2i\pi}}(z-w)^2
G_L(z)\lbrace{-}{1\over{{(z-u)^2}}}({\vec{A}}{\vec{\psi}})e^{ik_m{X^m}}
-{1\over{{(z-u)}}}\partial(({\vec{A}}{\vec{\psi}})e^{ik_m{X^m}})\rbrace\cr
=\oint{du}\oint{{dz}\over{2i\pi}}
-{{(z-w)^2}\over{{(u-w)^2}}}G_L(u)({\vec{A}}{\vec{\psi}})e^{ik_m{X^m}}(u)
-{{(z-w)^2}\over{{(u-w)}}}\partial(G_L({\vec{A}}{\vec{\psi}})e^{ik_m{X^m}}(u))
\cr
=\oint{du}\lbrace{2}(u-w)G_L({\vec{A}}{\vec{\psi}})e^{ik_m{X^m}}(u)
+(u-w)^2\partial(G_L({\vec{A}}{\vec{\psi}})e^{ik_m{X^m}}(u))
\cr=\oint{du}\partial((u-z)^2G_L(u)({\vec{A}}{\vec{\psi}})e^{ik_m{X^m}}(u))
=0}}.

Finally, for the sake of completeness, we also note that
 the commutator of $V^{\alpha-}$ at the $H_{-3}$ representation
 with the photon at picture $-1$ (and subsequently
all the pictures below)
produces the BRST trivial result:

\eqn\grav{\eqalign{{\lbrack}L^{\alpha-}_{(-3)},V_{ph}^{(-1)}\rbrack
=:\Gamma^{-2}{\lbrack}Q_{brst},\oint{du}(z-u)^2
{\partial\xi}G_L(u)({\vec{A}}{\vec{\psi}})e^{ik_m{X^m}}(u)\rbrack:}}

This constitutes the proof of the annihilation of $V_{ph}$ 
by $L^{\alpha-}$.

On the contrary, the remaining eight scalar $\alpha$-generators
acting on $V_{ph}$ produce new physical states
of $H_n\sim{H_{-n-2}},n=1,2,3$.
Thus the $H_{1}\sim{H_{-3}}$ $\alpha$-generator $L^{\alpha+}$
acts on the piture zero photon, producing the vertex operator  $V^{\alpha+}$,
the massless  gauge vector boson of the same ghost cohomology:
\eqn\grav{\eqalign{{\lbrack}L^{\alpha+}_{(-3)},V_{ph}^{(0)}(k)\rbrace
\cr=3\oint{du}e^{-3\phi}\lbrace
({\vec{k}}\partial{\vec{X}})({\vec{A}}\partial{\vec{X}})
({\vec{k}}{\vec{\psi}})+
({\vec{k}}{\vec{\psi}})({\vec{A}}{\vec{\psi}})
({\vec{k}}\partial{\vec{\psi}})
-({\vec{k}}\partial{\vec{X}})^2
({\vec{A}}{\vec{\psi}})\rbrace{e^{i{\vec{k}}{\vec{X}}}}}}
This operator is the element of $H_{-3}$. Its $H_1$ version is straightforward
to obtain by replacing $-3\phi\rightarrow\phi$ and adding
the $b-c$ correction term using the prescription of ~{\ghost}.
For the sake of completeness, 
 the explicit expressions for 
some of the remaining gluon operators are given below:
\eqn\grav{\eqalign{V^{\beta{+}}=\lbrack{L^{\beta+}},V_{ph}\rbrack\cr
=\oint{e^{-4\phi}}\lbrace\lambda\partial^2\varphi-2\partial\varphi
\partial\lambda\rbrace
\lbrace(-({\vec{k}}\partial{\vec{X}})^2+i({\vec{A}}\partial^2{\vec{X}})
({\vec{k}}{\vec{\psi}}))e^{i{\vec{k}}{\vec{X}}}-i({\vec{k}}{\vec{X}})\partial
W_{ph}\rbrace\cr
V^{\beta-}=\lbrack{L^{\beta-}},V_{ph}\rbrack
=\oint{e^{-4\phi}}
\lbrace{2}\lambda\partial\varphi\lbrack
(({\vec{k}}\partial{\vec{X}})^2
({\vec{A}}\partial{\vec{\psi}})
+({\vec{k}}\partial^2{\vec{X}})({\vec{k}}\partial{\vec{X}})
({\vec{A}}{\vec{\psi}})
\cr
-i({\vec{k}}\partial{\vec{\psi}})
({\vec{A}}\partial^2{\vec{X}})+
({\vec{k}}{\vec{\psi}})({\vec{k}}\partial{\vec{X}})
({\vec{A}}\partial^2{\vec{X}}))e^{i{\vec{k}}{\vec{X}}}
-{i\over2}({\vec{k}}\partial{\vec{\psi}})\partial{W_{ph}}(k)\rbrack\cr
-4\partial\lambda\partial\varphi
\lbrack
(({\vec{k}}{\vec{\psi}})({\vec{k}}\partial{\vec{X}})
({\vec{A}}\partial{\vec{X}})
-i({\vec{k}}\partial{\vec{\psi}})({\vec{A}}\partial{\vec{X}})
+
({\vec{k}}\partial{\vec{X}})^2({\vec{A}}{\vec{\psi}}))e^{i{\vec{k}}{\vec{X}}}
-i({\vec{k}}\partial{\vec{\psi}})W_{ph}(k)\rbrack\cr
-\lambda\partial^2\varphi\lbrack
(-({\vec{A}}{\vec{\psi}})({\vec{k}}\partial{\vec{X}})^2
+i({\vec{k}}{\vec{\psi}})({\vec{A}}\partial^2{\vec{X}}))e^{i{\vec{k}}{\vec{X}}}
+i({\vec{k}}{\vec{\psi}})\partial{W_{ph}}(k)\rbrack\rbrace\cr
V^{\alpha\beta}=\lbrack{L^{\alpha\beta}},V_{ph}\rbrack=
\oint{e^{-4\phi}}\lbrace({\partial}{L_1}(\varphi,\lambda)-2\partial
\phi{L_1}(\varphi,\lambda)\rbrace({\vec{k}}{\vec{\psi}})({\vec{k}}
\partial{\vec{X}})
({\vec{A}}{\vec{\psi}}))e^{i{\vec{k}}{\vec{X}}}\cr
V^{\beta\gamma}=\lbrack{L^{\beta\gamma}},V_{ph}\rbrack
=\oint{e^{-5\phi}}\lbrace
\partial{L_2}(\varphi,\lambda)-{{11}\over3}\partial\phi{L_2}(\varphi,\lambda)
\rbrace({\vec{k}}{\vec{\psi}})({\vec{k}}\partial{\vec{X}})
({\vec{A}}{\vec{\psi}})e^{i{\vec{k}}{\vec{X}}}\cr
V^{\gamma+}=\lbrack{L^{\gamma+}},V_{ph}\rbrack
=\oint{e^{-5\phi}}\lbrace\partial{F_2}(\lambda,\varphi)
-{{10}\over3}\partial\phi
F_2(\varphi,\lambda)\rbrace\cr\times
\lbrace(-({\vec{k}}\partial{\vec{X}})^2({\vec{A}}{\vec{\psi}})
+i({\vec{A}}\partial^2{\vec{X}})({\vec{k}}{\vec{\psi}}))e^{i{\vec{k}}{\vec{X}}}
-i({\vec{k}}{\vec{\psi}})\partial{W_{ph}}(k)\rbrace
}}
where $W_{ph}(k)=({\vec{A}}\partial{\vec{X}}+i({\vec{k}}{\vec{\psi}})
({\vec{A}}{\vec{\psi}}))e^{i{\vec{k}}{\vec{X}}}$
is the integrand of the photon vertex operator at picture zero.
The explicit expressions for the vertex operators
$V^{\gamma-}$ and $V^{\alpha\gamma}$
are quite lengthy (due to more complicated structure of the appropriate
$\alpha$-generators) and we will skip them here for the sake of brevity,
as we won't use them in any further calculations in this paper.
These expressions, if needed, are straightforward to obtain
by applying $L^{\gamma-}$ and $L^{\alpha\gamma}$ to the photon,
exactly as demonstrated above.
The operators (22) are given in the negative cohomology representations,
i.e. the elements of negative ghost cohomologies $H_{-n-2} (n=1,2,3)$
Their versions in the corresponding isomorphic 
positive cohomology representations
$H_n(n=1,2,3)$ are straightforward to obtain either by the direct
isomorphism construction described in ~{\ghost} or
(much easier) by replacing $(-n-2)\phi\rightarrow{n}$
and adding the correction terms using the formalism described 
ibide. In particular, some examples
of manifest expressions for positively represented
gluons will be considered in Section 5.
 In the following sections we will discuss the OPE properties and
the structure constants of the gluon vertex operators (22), (23)
and analyze their scattering amplitudes.

\centerline{\bf 4. Gluon Vertices:
Structure Constants and Closeness of the Operator Algebra}

In this section we analyze the structure constants
(3-point functions) of the operators (22), (23).
In particular, the remarkable property of the operators (22), (23), which
we will prove below,
is the closeness of their operator algebra, 
related to  their  underlying zigzag symmetry, reflecting the relevance 
of these operators to the gauge theory dynamics and gauge-string 
correspondence.
Using the explicit expressions for the operators (23), it is in principle
straightforward to calculate their structure constants.
There is no need, however, in the direct calculation
(not hard but somewhat long), as the form of the constants
is actually fixed by the $\alpha$-symmetry.
 Consider the 3-point correlation function of the
 standard open string photons:
\eqn\grav{\eqalign{<V_{ph}(k)V_{ph}(p)V_{ph}(q)>
= A_l(k)A_m(p)A_n(q)<cW_l(z_1)cW_m(z_2)cW_n(z_3)>\cr
=i\lbrace({\vec{k}}{\vec{A}}({\vec{q}}))({\vec{A}}({\vec{k}}){\vec{A}}
({\vec{p}}))
-({\vec{q}}{\vec{A}}({\vec{p}}))({\vec{A}}({\vec{k}}){\vec{A}}({\vec{q}}))
+({\vec{p}}{\vec{A}}({\vec{k}}))({\vec{A}}({\vec{p}}){\vec{A}}({\vec{q}}))}}
where, for the  certainty, $V_{ph}$ taken in unintegrated form and
 $W_m$ are the dimension 1 integrands of the integrated photon operator:
$W_n=e^{-\phi}{\psi_n}e^{i{\vec{k}}{\vec{X}}}$ at picture $-1$
and $W_n=(\partial{X_n}+i({\vec{k}}{\vec{\psi}})\psi_n)e^{i{\vec{k}}{\vec{X}}}$
at picture 0 (note that
the extra term $\sim\gamma\psi_n{e^{i{\vec{k}}{\vec{X}}}}$
in the picture zero expression for the unintegrated photon
is ignored here since it doesn't contribute to the 3-point function;
two photon operators in the correlator (24)
 must be taken at picture $-1$ and the third one at picture 0).
With the correlation function (24), the OPE of two dimension 1 
photon operators 
is 
\eqn\grav{\eqalign{W_l(k;z_1)W_m(p;z_2)=
{{C_{lm}^n(k,p)W_n(q;{{z_1+z_2}\over2})}\over{z_1-z_2}}\cr
+\sum_{N=0}^\infty(z_1-z_2)^{N+{{({\vec{q}})^2}\over2}}
{C^{(N)}}(k,p){W^{(N)}}(q)\cr
C_{lm}^n(k,p)=i(k^n\eta_{lm}-q_m\eta_l^n+p_l\eta_m^n)\cr
{\vec{k}}+{\vec{p}}+{\vec{q}}=0}}
where we have skipped BRST trivial tachyonic term of the  order of
$(z-w)^{-2}$.
The $W^{(N)}(q)$-operators
appearing in the higher order terms of the OPE (25) correspond to the massive
 poles in the Veneziano amplitude (after the appropriate on-shell conditions
on q are imposed)
Structurally,  these operators have the form
\eqn\lowen{W^{(N)}(q)=U_{N+2}(X,\psi)e^{iq{X}}}
where $U_{N+2}$ is some polynomial in $X$ and $\psi$ of conformal
dimension $N+2$ and the on-shell condition for $W^{(N)}$
(corresponding to the mass of the appropropriate physical pole in the 
Veneziano amplitude) is given by
\eqn\lowen{({\vec{q}})^2=-2N-2}

Note that, apart from the color related factor, the structure constants
$C_{lm}^n$ in front of the photon 
simply reproduce the $3$-gluon vertex of QCD.
 The next step is to  apply $\alpha$-transform
to the both sides of the OPE (25).
For the brevity, it is convenient to write 
\eqn\grav{\eqalign{
\lbrack{L^i},{L^j}\rbrack=D^{ij}_k{L^k} (i=1,...,8); \cr
 V_m^i\equiv\lbrack{L^i},W_m\rbrack}}

where $L^i$ stands for any of eight SU(3) $\alpha$-generators
(17) - $F_\ph,G_\pm,F_3,G_3$ and $L_{1,2}$ , $V^i$ are the corresponding gluon
vertices
and $T^{ij}_k$ are the SU(3) structure constants.
Consider the OPE of two gluon integrands of dimension 1:
\eqn\grav{\eqalign{W_l^i(k;z_1)W_m^j(p;z_2)
=\oint_{z_1}{{dw_1}\over{2i\pi}}\oint_{z_2}{{dw_2}\over{2i\pi}}
T^i(w_1)T^j(w_2)W_l(k;z_1)W_m(p;z_2)}}
with $k$ and $p$ being the on-shell momenta of 2 photons
and $T^i,T^j$ are the integrands of $L_i,L_j$.
Firstly, let's concentrate on the simple pole of the OPE.
Since $T^i$ and $T^j$ are the dimension one 
primaries and the physical operators,
the only BRST non-trivial operator in their full OPE will be that 
of the dimension one, i.e. the one appearing in the simple pole term
$\sim(z_1-z_2)^{-1}$. This operator is simply the integrand of the r.h.s.
of the commutator of $L_i$ and $L_j$, i.e.
$D^{ij}_k{T^k}(z)$ where $T^k$ is the integrand of $L^k$.
Therefore the simple pole in the OPE  of $W_i^l$ and $W^j_m$ is given by
\eqn\grav{\eqalign{W_l^i(k,z_1)W_m^j(p,z_2)
\sim{{D^{ij}_{k}C_{lm}^n(k,p,q)}\over{z_1-z_2}}
\oint_{{{z_1+z_2}\over2}}
{{dw}\over{2i\pi}}{\lbrack}T^k(w),W_n(q;{{z_1+z_2}\over2})\rbrack\cr
={1\over{z_1-z_2}}{{D^{ij}_{k}}C_{lm}^n(k,p,q)}W_n^k(q;{{z_1+z_2}\over2})}}
in particular, this indicates
that the 3-point gluon vertex is reproduced by the OPE (30).
But what about the full OPE of two gluons, that is,
 the higher order terms ? In the standard case of two photons, these terms
(with the on-shell condition (27) imposed) give rise to the massive
intermediate states in any $n$-point amplitude with $n\geq{4}$
, killing the ``naive'' gauge-string correspondence.
How about gluons? Remarkably, it turns out that all 
the intermediate massive states in the OPE of two gluons (30)
are BRST-trivial;  for this reason they do not
lead to any massive poles (destroying the zigzag invariance) and
thus the gauge-string correspondence is protected.
This occurs due to very special property of the 
SU(3) $\alpha$-generators
(proof given below): they produce new physical states only
when applied to massless vertex operators; however, the $\alpha$-transform of
any massive superstring mode gives BRST-trivial result!
Here is the proof. Repeating the arguments we used to fix the 
3-gluon vertex (30) and using the full OPE (25) of two
photons, it is easy to check  that the term of the order
of $(z_1-z_2)^{n}$  in the OPE of two gluons $W_m^i(k)$ and $W_m^j(p)$
is given by $\sim{C^{(N)}}(k,p,q)D^{ij}_k
\lbrack{L^k},{W^{(N)}}(q)\rbrack$
Our goal is thus to show that the operator
$\lbrack{L^k},{W^{(N)}}(q)\rbrack$ is BRST-trivial for any 
$({\vec{q}})^2\neq{0}$.
For brevity,below we will give the proof for the case
$L^{(k)}={L^{\alpha+}}$; the proof is
totally
analogous for all other SU(3) $\alpha$-generators.

 To simplify things further, it is more convenient
to take
the  massive operators of the right-hand side of (25)
at the unintegrated $b-c$-picture, that is,
\eqn\lowen{V^{(N)}(q)=cW^{(N)}(q)=cU_{N+2}(X,\psi)e^{i{\vec{q}}{\vec{X}}}}
(since the unintegrated and integrated vertices are related by
the BRST-invariant $Z$-transformation ~{\sellf, \ghost}, any 
(non)triviality statement proven for an unintegrated vertex,
applies to the integrated one as well)
The BRST-invariance condition for the vertex operator (31)
implies
\eqn\lowen{\lbrace{Q_{brst}},cU_{N+2}\rbrace
e^{i{\vec{q}}{\vec{X}}}=cU_{N+2}\lbrack{Q_{brst}},e^{i{\vec{q}}{\vec{X}}}
\rbrack.}
where
\eqn\lowen{\lbrack{Q_{brst}},e^{i{\vec{q}}{\vec{X}}}
\rbrack= {1\over2}({\vec{q}})^2{\partial{c}}e^{i{\vec{q}}{\vec{X}}}
+c\partial(e^{i{\vec{q}}{\vec{X}}})+{i\over2}\gamma({\vec{k}}{\vec{\psi}})
e^{i{\vec{q}}{\vec{X}}}}
Writing
\eqn\grav{\eqalign{U^{\alpha+}_{N+2}\equiv\lbrack{L^{\alpha+}},
U_{N+2}\rbrack\cr
\lbrack{L^{\alpha+}},e^{i{\vec{q}}{\vec{X}}}\rbrack=Z^{\alpha+}
{e^{i{\vec{q}}{\vec{X}}}}\cr
V^{\alpha+}_{N+2}(q)\equiv{\lbrack}L^{\alpha+},V_{N+2}(q)\rbrack\cr
=(U_{N+2}^{\alpha+}+U_{N+2}Z^{\alpha+})e^{i{\vec{q}}{\vec{X}}}
}}
where
\eqn\lowen{Z_{\alpha+}=i\partial(e^{-3\phi})({\vec{k}}{\vec{\psi}})+3
i{e^{-3\phi}}({\vec{k}}\partial{\vec{\psi}}),}
we apply the $\alpha$-transformation by $L^{\alpha+}$ to the both sides of 
(33)
obtaining
\eqn\grav{\eqalign{
\lbrace{Q_{brst}},cU^{\alpha+}_{N+2}\rbrace{e^{i{\vec{q}}{\vec{X}}}}
+\lbrace{Q_{brst}},cU_{N+2}\rbrace{Z^{\alpha+}}{e^{i{\vec{q}}{\vec{X}}}}
\cr
={1\over2}({\vec{q}})^2
c\partial{c}U^{\alpha+}_{N+2}{e^{i{\vec{q}}{\vec{X}}}}
+{1\over2}({\vec{q}})^2{c}\partial{c}U_{N+2}Z^{\alpha+}
{e^{i{\vec{q}}{\vec{X}}}}
=-{1\over2}({\vec{q}})^2\partial{c}V^{\alpha+}_{N}(q)
}}
where we used the commutation relations (34), as well as
\eqn\lowen{
:\gamma{Z^{\alpha+}}:=:\gamma{e^{-3\phi}}:=:c^2:=0.}
Substituting the identity (32) we cast the equation (36) as
\eqn\grav{\eqalign{
\lbrace{Q_{brst}},cU^{\alpha+}_{N+2}\rbrace{e^{i{\vec{q}}{\vec{X}}}}
+cU_{N+2}Z^{\alpha+}\lbrack{Q_{brst}},e^{i{\vec{q}}{\vec{X}}}\rbrack
=
-{1\over2}({\vec{q}})^2\partial{c}V^{\alpha+}_{N}(q).}}
But
\eqn\lowen{\lbrace{Q_{brst}},cU^{\alpha+}_{N+2}\rbrace
=-\lbrace{Q_{brst}},cU_{N+2}Z^{\alpha+}\rbrace}
since the operator $:U^{\alpha+}_{N+2}+U_{N+2}Z^{\alpha+}:$ is BRST-invariant.
In fact, this point deserves  separate clarification.
That is, the BRST-invariance
 of  $:U^{\alpha+}_{N+2}+U_{N+2}Z^{\alpha+}:$
 is guaranteed since the structure of this operator is
inherited from the  appropriate term of the $\alpha$-transformed OPE 
(25) of two 
photons.
  Since the photons of the OPE (25) are on-shell, the BRST commutator
with each term in their full
OPE has to vanish separately.
If the $\alpha$-transformed term of the order of $(z_1-z_2)^N$
in the OPE (25) is given by ${C^{(N)}}\lbrack{L^{\alpha+}},W^{(N)}(q)\rbrack$,
the 
corresponding operator in the OPE of two gluons is given by
$cW^{{\alpha+},{(N)}}(q)=\lbrack{L^{\alpha+}},cW^{(N)}(q)\rbrack=
c(U^{\alpha+}_{N+2}+U_{N+2}Z^{\alpha+})
{e^{i{\vec{q}}{\vec{X}}}}$ at the unintegrated $b-c$ picture.
This operator is invariant for all values $q$, both on and off-shell
 (as a matter of fact, it is exact
for all  $({\vec{q}})^2\neq{2N+2}$).
In the special off-shell case $({\vec{q}})^2=0$
using 
 (37) along with the commutator (33)
gives
\eqn\grav{\eqalign{0=\lbrace{Q_{brst}},cW^{{\alpha+},{(N)}}(q)\rbrace
=\lbrace{Q_{brst}},c(U^{\alpha+}_{N+2}+U_{N+2}Z^{\alpha+})\rbrace
{e^{i{\vec{q}}{\vec{X}}}}\cr{-}
c(U^{\alpha+}_{N+2}+U_{N+2}Z^{\alpha+})\lbrack
Q_{brst},{e^{i{\vec{q}}{\vec{X}}}}\rbrack
=\lbrace{Q_{brst}},c(U^{\alpha+}_{N+2}+U_{N+2}Z^{\alpha+})\rbrace
{e^{i{\vec{q}}{\vec{X}}}}}}
as the second term in the commutator (40) vanishes due to (33) and (37).
Thus  $:U^{\alpha+}_{N+2}+U_{N+2}Z^{\alpha+}:$ is BRST-invariant and therefore,
in view of (40), the identity (36) can be written as

\eqn\grav{\eqalign{-\lbrace{Q_{brst},U_{N+2}Z^{\alpha+}}\rbrace
{e^{i{\vec{q}}{\vec{X}}}}
+cU_{N+2}Z^{\alpha+}\lbrack{Q_{brst}},{e^{i{\vec{q}}{\vec{X}}}}\rbrack
\cr
\equiv
-\lbrace{Q_{brst}},cU_{N+2}Z^{\alpha+}{e^{i{\vec{q}}{\vec{X}}}}\rbrace
=-{1\over2}({\vec{q}})^2\partial{c}V^{\alpha+}_{N}(q)}}
Now, since the unintegrated vertex operator $V^{\alpha+}_{N}(q)$
is dimension zero primary field, it is annihilated by
 the zero mode of the full stress tensor:
$\lbrack{T_0},V^{\alpha+}_{N}(q)\rbrack=0$ where
$T_0={\oint}{{dz}\over{2i\pi}}zT(z)$.  Therefore, as
$T_0=\lbrace{Q_{brst}},b_0\rbrace$, the identity (41) implies:
\eqn\lowen{V^{\alpha+}_N(q)={{2\over{({\vec{q}})^2}}}
\lbrack{Q_{brst}},b_0cU_{N+2}Z^{\alpha+}e^{i{\vec{q}}{\vec{X}}}\rbrack}
This concludes the proof that the $\alpha$-transformation of
any massive physical operator by $L^{\alpha+}$ is BRST-exact.
The proof is completely analogous for any other 
$\alpha$-generator $L^k;k=1,...,8$, so the generalization of (42)
for an arbitrary $\alpha$-generator transforming a massive
physical operator, is
\eqn\grav{\eqalign{
V^{k}_{N}(q)\equiv\lbrack{L^k},V_N(q)\rbrack
={{2\over{({\vec{q}})^2}}}
\lbrack{Q_{brst}},b_0cU_{N+2}Z^{k}e^{i{\vec{q}}{\vec{X}}}\rbrack
}}
with the operators $Z^k(q)$ defined according to
\eqn\lowen{\lbrack{L^k},e^{i{\vec{q}}{\vec{X}}}\rbrack
=Z^k(q)e^{i{\vec{q}}{\vec{X}}}}
This constitutes the proof of the closeness (and, accordingly,
 of the underlying zigzag symmetry) of the gluon operator algebra (30).
Speaking metaphorically, the $\alpha$-symmetry transform, applied
to the OPE of photons, acts like an Occam's razor: it shaves off
 the infinite tower of the higher order on the right-hand side
(leading to massive poles in scattering amplitudes) protecting
the  zigzag symmetry of the photons turned gluons.
In the following sections we will use this
 ``lex parsimoniae'' to directly compute
 the 4-point
scattering amplitude involving the gluons.

\centerline{\bf 5. Zigzag-Invariant vs. Veneziano amplitudes: 
technical  remarks}

Using the results of the previous section (zigzag symmetry of 
the constructed gluon vertex operators and the closeness of their OPE (30))
it is now relatively easy to calculate 
N-point scattering amplitudes
of gluons with higher number of vertices ($N>3$) from string theory,
despite a complicated formal structure of the vertex operators
(22),(23).

Since all the massive intermediate states 
appearing in the operator product of any two gluon operators are BRST-trivial,
they do not contribute to worldsheet correlations,
so only the massless gluon vertices 
appearing on the right hand side of the OPE
(multiplied by the 3-gluon structure constants) are relevant.
Therefore we can use
the simplicity of the zigzag-invariant operator algebra 
of gluons to compute their correlation functions by bootstrap.
This calculation (which can be generalized to higher number
of scattering gluons as well as to include the loop corrections)
will be  performed in the section 6 of this paper.
Before we proceed with the calculations, however, it is instructive 
to make few technical 
comments concerning the structure of the $4$-point amplitude.

As our principal  calculation  of the 4-point 
amplitude is almost independent 
on the arguments of this section, a reader, not interested
in the technical details, can skip it and go directly to the Section 6,
where this calculation is presented.

As we already noted above, we expect the 4-point amplitude
of gluon vertex operators (22),(23) to have  field-theoretic behaviour,
implying the absence of massive resonances.
Such a  behaviour  immediately 
follows from with the OPE (30) for the gluons and the 
underlying zigzag symmetry;
 however,
it may seem a bit of a surprise from technical point of view.
Indeed, the structure of  4-point 
Veneziano amplitude in ``orthodox''
string theory is well-known; it has
an infinite sequence of massive poles, corresponding to 
integer values of the Mandelstam parameters.
The question is - what is so special about the gluon operators (22),(23)
that distinguishes them from usual vertex operators (such as photons) 
leading to radically different pole structure?
Not quite surprisingly, the answer lies in the ghost number 
structure of the gluon operators and their ``non-standard'' ghost numbers.
To answer the question, it is useful to recall  how the ``standard''
Veneziano amplitude emergees in open string theory.
Typically, the 4-point amplitude of open string theory
involves $3$ unintegrated vertex operators and one integrated, i.e.
it has the form $\sim{\int_0^1{dz}}<W(z)cW_1(z_1)cW_2(z_2)cW_3(z_3)>$
with the points $z_1,z_2,z_3$ fixed and the remaining operator $W(z)$
integrated over the worldsheet boundary. Such a structure is dictated by the
$b-c$ ghost number anomaly which is equal to $-3$, so
to cancel it, one needs three
unintegrated vertices, each of them carrying the $b-c$ ghost number $+1$.
Here the $b-c$ ghost factor $<c(z_1)c(z_2)c(z_3)>$ particularly leads
to the standard Koba-Nielsen measure 
given by $\sim(z_1-z_2)(z_1-z_3)(z_2-z_3)$. Using the $SL(2,C)$
invariance, one fixes $z_1=0,z_2=1,z_3=\infty$ so the 
ermaining $z$ integral is proportional to $\sim{\int_0^1}dz
z^{-{s\over2}}(1-z)^{-{t\over2}}$ which is just the Euler's beta-function
of the Mandelstam variables, leading to the Veneziano amplitude.
Such is the standard case situation. With the gluon operators
(22), (23) things are more subtle. Since these operators 
are the elements of nonzero ghost cohomologies, they do not exist at
arbitrary ghost numbers and one has to be careful to 
ensure the correct balance of ghost numbers to cancel both
superconformal ghost number anomalies (equal to $+2$ for the $\phi$-field 
and $-1$ for the $\chi$-field), in addition to the $b-c$ anomaly.
And this is where the difference strikes.

Consider the $4$-point functions of the gluon operators (22), (23)
with two of them being in the positive
and two in negative ghost cohomology representations
(for the reasons explained in ~{\ghost}
any correlation function involving the picture-dependent operators 
always has to involve the operators from cohomologies of opposite signs).
The peculiar property of operators 
in  positive cohomologies is that they only exist 
in the integrated form, as the $b-c$
correction terms that ensure the overall BRST
invariance of the positive cohomology elements
can only be constructed in the integrated case ~{\ghost}. 
 This is related to the fact that integrated
and unintegrated forms of vertex operators correspond to their
representations in two adjacent $b-c$ pictures, which are
the fermionic analogues of the usual superconformal ghost pictures,
with the BRST-invariant $Z$-transformation being the analogue 
of picture-changing operator $\Gamma$.
~{\sellf, \ghost}
 Not surprisingly, the inequivalence in the superconformal pictures
(resulting in the emergence of ghost cohomologies) is related
to inequivalence in the fermionic  
$b-c$-pictures, in view of the supersymmetry.
In fact, there is an underlying geometrical principle relating the 
bosonic and fermionic ghost cohomologies, which is not yet understood
completely and,  by itself, it is an interesting direction
 for the future research.
Therefore, contrary to the case of  picture-independent generators
which 4-point function contains only one integrated vertex,
leading to ``standard'' Veneziano
amplitude, the 4-point function of gluons has to contain
at least  2 integrated
operators, corresponding to insertions from positive cohomologies.
 At the  same time, the presence of two integrated vertices
in the 4-point function does not violate the $b-c$ ghost number balance
since, as we know, the integrated operators of positive cohomologies
involve the correction terms proportional to the $c$-ghost field.
These correction terms now play a significant role; it is
because of them that one can have a $4$-point correlator 
with two integrated vertices
but with the total $b-c$ ghost number of the correlator
still equal to 3 (needed to cancel the $b-c$-anomaly).
This  modifies the standard Koba-Nielsen's determinant,
leading to very different overall result for the scattering amplitude.
To be more concrete, consider an example of the 4-gluon amplitude:
$<V^{\beta\gamma}(p_1)V^{\gamma+}(p_2)V^{\alpha\beta}(p_3)V^{\alpha+}(p_4)>$.
To ensure the ghost number anomaly cancellation,
we take $V^{\beta\gamma}$ and $V^{\gamma+}$
(the elements of $H_{3}\sim{H_{-5}}$) in the positive representation
while $V^{\alpha\beta}$ and $V^{\alpha+}$
(the elements of $H_{2}\sim{H_{-4}}$ $H_{1}\sim{H_{-3}}$ in the negative.
The negatively represented operators are taken in the
unintegrated form and are
 given by the integrands of the appropriate expressions
of (22), (23) multiplied by $c$, while
the positively represented operators of $H_3$, including
the necessary $b-c$ correction terms are

\eqn\grav{\eqalign{
V^{\beta\gamma}=\lbrack{L^{\beta\gamma}},V_{ph}\rbrack
=\oint{dz}{e^{3\phi}}\lbrace
\partial{L_2}(\varphi,\lambda)-{{11}\over3}\partial\phi{L_2}(\varphi,\lambda)
\rbrace({\vec{k}}{\vec{\psi}})({\vec{k}}\partial{\vec{X}})
({\vec{A}}{\vec{\psi}})e^{i{\vec{k}}{\vec{X}}}(z)\cr
-{{22}\over{15}}\oint_u{dz}(u-z)^5{\lbrace}{c}e^{2\phi+\chi}
P^{(3)}_{\phi-\chi}
({\vec{k}}{\vec{\psi}})({\vec{k}}\partial{\vec{X}})
({\vec{A}}{\vec{\psi}})e^{i{\vec{k}}{\vec{X}}}(z)
+e^{3\phi}G^{(1)}_{{{27}\over2}}(L_2,X,\psi,\phi,\chi,\sigma)\rbrace\cr
V^{\gamma+}=\lbrack{L^{\gamma+}},V_{ph}\rbrack
=\oint{dz}{e^{3\phi}}\lbrace(\partial{F_2}(\lambda,\varphi)
-{{10}\over3}\partial\phi
F_2(\varphi,\lambda))(z)\cr\times
((-({\vec{k}}\partial{\vec{X}})^2({\vec{A}}{\vec{\psi}})
+i({\vec{A}}\partial^2{\vec{X}})({\vec{k}}{\vec{\psi}}))e^{i{\vec{k}}{\vec{X}}}
-i({\vec{k}}{\vec{\psi}})\partial{W_{ph}}(z,{\vec{k}}))(z)\rbrace\cr
-{2\over3}\oint_u{dz}(u-z)^5\lbrace
\cr
{c}e^{2\phi+\chi}P^{(3)}_{\phi-\chi}
L_2(\varphi,\lambda)(z)(
-({\vec{k}}\partial{\vec{X}})^2({\vec{A}}{\vec{\psi}})
+({\vec{k}}{\vec{\psi}})({\vec{k}}\partial
{\vec{\psi}})
({\vec{A}}{\vec{\psi}})+({\vec{A}}\partial{\vec{X}})({\vec{k}}\partial
{\vec{X}})({\vec{k}}{\vec{\psi}}))(z)\cr
+e^{3\phi}G^{(2)}_{{{27}\over2}}(F_2,X,\psi,\phi,\chi,\sigma)\rbrace
}}
where the $z$-integrals in the correction terms are
taken around some point $u$ of the worldsheet boundary
(which can be fixed to zero by the $SL(2,C)$ symmetry)
and $P^{(3)}_{\phi-\chi}$ is conformal dimension 3 polynomial
in the derivatives of $\phi$ and $\chi$ defined according to
$P_{L(\phi_1(z),...\phi_n(z))}^{(n)}=e^{-L(\phi_1(z),
...\phi_n(z))}{{d^n}\over{dz^n}}e^{L(\phi_1(z),...,\phi_n(z))}$
with $L$ being an arbitrary given function of 
arbitrary $n$ fields $\phi_1(z),...,\phi_n(z)$ 
($L=\phi-\chi$ in our case).
The conformal dimension ${{27}\over2}$ operators
$G^{(1)}_{{{27}\over2}}(L_2,X,\psi,\phi,\chi,\sigma)$
$G^{(2)}_{{{27}\over2}}(F_2,X,\psi,\phi,\chi,\sigma)$
are the polynomials in the worlsheet fields
 $F_2(\varphi,\lambda),L_2(\varphi,\lambda),X,\psi,\phi,\chi,\sigma$
and their derivatives.
They have complicated explicit form but fortunately
their manifest expressions are not needed for any calculations 
in this paper. These expressions can be obtained straightforwardly
by evaluating the fifth order non-singular terms in the operator products:
\eqn\grav{\eqalign{
:ce^{2\chi-2\phi}:(z):{b}e^{5\phi-2\chi}\lbrace
{1\over{120}}P^{(5)}_{2\phi-2\chi-\sigma}(\partial{L_2}-{{11}\over3}
\partial\phi{L_2})
\cr
-{{11}\over{1080}}P^{(6)}_{{2\phi-2\chi-\sigma}}{L_2}\rbrace
R_1(X,\psi):(w)\cr
\sim{...}+(z-w)^5
:{e^{3\phi}}G^{(1)}_{{{27}\over2}}:({{z+w}\over2})+...\cr
:ce^{2\chi-2\phi}:(z) :{b}e^{5\phi-2\chi}\lbrace{1\over{120}}
{P^{(5)}_{2\phi-2\chi-\sigma}}(\partial{F_2}-{{10}\over3}\partial\phi{F_2})
\cr
-{1\over{108}}P^{(6)}_{2\phi-2\chi-\sigma}{\rbrace}R_2(X,\psi):(w)\cr
\sim{...}+(z-w)^5:{e^{3\phi}}G^{(2)}_{{{27}\over2}}:({{z+w}\over2})+...
}}

where 
$$R_1(X,\psi)=({\vec{k}}{\vec{\psi}})({\vec{k}}\partial{\vec{X}})
({\vec{A}}{\vec{\psi}})e^{i{\vec{k}}{\vec{X}}}(z)$$
and
$$R_2(X,\psi)=-({\vec{k}}\partial{\vec{X}})^2({\vec{A}}{\vec{\psi}})
+i({\vec{A}}\partial^2{\vec{X}})({\vec{k}}{\vec{\psi}})e^{i{\vec{k}}{\vec{X}}}
-i({\vec{k}}{\vec{\psi}})\partial{W_{ph}}(z,{\vec{k}})$$

The ghost number anomaly cancellation condition then 
determines that the 4-point
correlator is given by the crossterm contribution
of the  basic term (proportional to $e^{3\phi}$ )
in one 
of two positively represented operators ($V^{\gamma+}$ or $V^{\beta\gamma}$)
and the correction term in the second (proportional to $ce^{2\phi+\chi}$)
, as the ghost factors of  the negatively represented 
unintegrated operators,
$V^{\alpha\beta}(z_3)$ and $V^{\alpha+}(z_4)$ are given by
to $c{e^{-4\phi}}$ and ${c}e^{-3\phi}$ respectively
(with $z_{3,4}$ being the locations of the operators).
Using the $SL(2,C)$ symmetry, we can fix
$z_1=0,z_3=1,z_4=\infty$, so the resulting double integral
for the $4$-point function consists of the
terms with the structure
$$A\sim\int_0^1{dz}\int_0^1{du}{u^5}(z-u)^a(u-1)^b(z-1)^c$$
where $a,b$ are linear in the Mandelstam parameters.
(e.g. typically, $a(s)={s\over2}+m, b(t)
={t\over2}+n,c=-s-t+p$ where $m,n$ and $p$ are integer numbers
depending on the particular contraction) and the $u^5$-factor
originates from non-local $c$-dependent correction terms
of positively represented gluon vertices.

The integration in  $z$ can be done analytically if 
 $Re\lbrace{c}\rbrace\geq{-1}$ and the answer is given by
$$A\sim{(1+c)^{-1}u^{5+a}(u-1)^b{}_2F_{1}(1,-a,2+c;{{1}\over{u}})}.$$
The subsequent integration in $u$ is also possible, if
$Re{\lbrace{b}\rbrace}>1$
leading to long and cumbersome
 combination of 
terms with the structure 
\eqn\grav{\eqalign{A\sim\sum_{m_1=1}^6
{{P_{m_1}(a,b)}\over{(1+c)}}
{{{}_2F_1}(m_1,m_1+1-c,m_1+7+a+b;1)\over{{\Gamma(1-a)}\Gamma(a+b+13)}}
\cr
-{{120{\pi}Csc(\pi{a})\Gamma(2+c){}_2F_1(-6,-1-a-c,7+b;1)}\over
{(1+c){\prod_{j=1}^6(j+b)}\Gamma(-a)\Gamma(2+a+c)}}}}
where $m_1$ are integers running from 1 to 6,
$P_{m_1}(a,b)$ are polynomials in $a$ and $b$
(lengthy and different for each ${m_1}$)

The overall number of terms turns out to be annoyingly huge, as
 the already cumbersome manifest expressions for the 
gluon vertices (22), (23), (45) lead to hudge number of contraction
with each contraction itself producing
lengthy combination of terms with different values of $m,n,p$ and $m_1$.
Despite such a complicated full expression for the amplitude obtained
by the direct integration, its structure described above is already 
sufficient to  understand qualitatevely the absence of stringy 
pattern involving
 infinite number massive poles. For each given $b$ and $c$
(recall that the integration is performed for $Re\lbrace{b,c}\rbrace\geq{-1}$)
it is the presence of two
$\gamma$-functions in the denominator that screens off the massive 
poles corresponding to large
negative integer values of $a$, or the tachyonic poles
related to integer positive $a$ values, if $|a|$ becomes significantly
larger
that $m_1$.

Fortunately 
the cumbersome expression (47)
emerging as a result of $z$ and $u$ integrations simplifies
radically in the on-shell limit
and the bootstrap calculation, demonstrated in the next section
indicates that in the end all this multitude of terms 
must conspire to converge to
quite an elegant answer with only the massless pole remaining.
Nevertheless the expression (47) is still of some interest as it 
 illustrates (qualitatively at least)
how the non-standard ghost coupling of the gluon vertices
(elements of nonzero ghost cohomologies) modifies
the Venezino amplitude, reducing its strusture from stringy 
to field-theoretic and truncating the infinite tower of the
massive modes.
In the next section we will demonstrate
the direct computation of the 4-point gluon amplitude,
based on the zigzag invariance of the OPE (30).

\centerline {\bf 6. Computation of the 4-point Amplitude}

In this section, we compute the 4-point gluon scattering amplitude:

\eqn\lowen{A^{i_1...i_4}({\vec{p_1}},
{\vec{p_2}}, {\vec{p_3}}, {\vec{p_4}})=<V^i_1({\vec{p_1}})V^{i_2}
({\vec{p_2}})
V^{i_3}({\vec{p_3}})V^{i_4}({\vec{p_4}})>}

using the closeness of the OPE (30).
We have

\eqn\grav{\eqalign{
A^{i_1...i_4}({\vec{p_1}},...,{\vec{p_4}})
={\prod_{j=1}^4}A^{m_j}({\vec{p_j}})\cr
<\oint_{z_1}T^{i_1}...\oint_{z_4}T^{i_4}
\int_0^1{dz_1}W^{m_1}(z_1;{\vec{p_1}})
\int_0^1{dz_2}W^{m_2}(z_2;{\vec{p_2}})cW^{m_3}(z_3;{\vec{p_3}})cW^{m_4}
(z_4,{\vec{p_4}})>}}

where $W^{m_i}$ are the dimension 1 integrands of photon vertex operators
and  the $z_a$ subscripts  in the contour integrals refer
to the points around which the contour integrals are taken
(two photon operators have to be at the integrated $b-c$ picture and
the remaining two are to be taken unintegrated, as we explained
 in the previous section). 
As previously, since $\oint{T^i}$ are the physical operators
of zero momentum, the only BRST non-trivial terms in the
full operator product of any two
$\oint{T^i}(z)$ and $\oint{T^j}(w)$  are those with conformal dimension 1,
i.e. of
the order $(z-w)^{-1}$, given by $D^{ij}_k\oint{T^k}$.
Furthermore, in the bootstrap of any two photon operators  (25)
it is sufficient to retain the massless terms since all the massive
operators become BRST-trivial after the appropriate $\alpha$-transform
by $\oint{T^k}$, as a consequence of the zigzag symmetry of the gluon
OPE (30).
To elucidate the pole structure of the amplitude (48),
it is convenient to take the photon operators slightly off-shell
first, imposing the on-shell condition
$({\vec{p_i}}{\vec{p_j}})=0;i=1,..,,4$ upon the calculation.

Then using (25) and (28), the 4-point correlator (49) is written as

\eqn\grav{\eqalign{{A^{i_1...i_4}}({\vec{p_1}},
{\vec{p_2}},{\vec{p_3}},{\vec{p_4}})
={\prod_{j=1}^4}A^{m_j}({\vec{p_j}})\int_0^1dz_1\int_0^1dz_2\lbrace
(z_1-z_2)^{(\vec{p_1}\vec{p_2})-1}C_{m_1m_2}^n
({\vec{p_1}},{\vec{p_2}})\cr
<R^{i_1...i_4}(z_1,..,z_4)W_n(z_2;{\vec{p_1}}+{\vec{p_2}})
cW_{m_3}(z_3;{\vec{p_3}})
cW_{m_4}(z_4,{\vec{p_4}})>
+
(z_1-z_3)^{(\vec{p_1}\vec{p_3})-1}\cr{\times}C_{m_1m_3}^n
({\vec{p_1}},{\vec{p_3}})
<R^{i_1...i_4}(z_1,..,z_4)W_{m_2}(z_2;{\vec{p_2}})cW_{n}(z_3;
{\vec{p_1}}+{\vec{p_3}})
cW_{m_4}(z_4; {\vec{p_4}})>
\cr+
(z_1-z_4)^{(\vec{p_1}\vec{p_4})-1}C_{m_1m_4}^n
({\vec{p_1}},{\vec{p_4}})\cr{\times}
<R^{i_1...i_4}(z_1,..,z_4)W_{m_2}(z_2;{\vec{p_2}})cW_{m_3}(z_3;
{\vec{p_3}})
cW_{n}(z_4;{\vec{p_1}}+{\vec{p_4}})>
\rbrace
}}
where the structure constants $C_{lm}^n({\vec{p_1}},{\vec{p_2}})$
 are given by the photon 3-vertex (25)

and the $\alpha$-generator insertion 
(related to the colour group theoretic factor)

is transformed to

\eqn\grav{\eqalign{R^{i_1...i_4}(z_1,...,z_4)=
(D^{i_1i_2}_j\oint_{z_2}T^j\oint_{z_3}T^{i_3}\oint_{z_4}T^{i_4}
+
D^{i_1i_3}_j\oint_{z_2}T^{i_2}\oint_{z_3}T^{j}\oint_{z_4}T^{i_4}
\cr
+
D^{i_1i_4}_j\oint_{z_2}T^{i_2}\oint_{z_3}T^{i_3}\oint_{z_4}T^{j}
+
D^{i_2i_3}_j\oint_{z_1}T^{i_1}\oint_{z_3}T^{j}\oint_{z_4}T^{i_4}
\cr
+
D^{i_2i_4}_j\oint_{z_1}T^{i_1}\oint_{z_3}T^{i_3}\oint_{z_4}T^{j}
+
D^{i_3i_4}_j\oint_{z_1}T^{i_1}\oint_{z_2}T^{i_2}\oint_{z_4}T^{i_j}
)}},
so 
we have expressed the $4$-point amplitude in terms of
$3$-point gluon vertices given  in
(30) 
Keeping in mind that the momenta are still slightly off-shell,
the three-point worldsheet correlators of gluon integrands (30)
are given by
\eqn\grav{\eqalign{<W^i_{n_1}(z_2;{\vec{k}})W^j_{n_2}(z_3;{\vec{p}})
W^k_{n_3}(z_4;{\vec{q}})>\cr
=(z_2-z_3)^{({\vec{k}}{\vec{p}})-1}(z_2-z_4)^{({\vec{k}}{\vec{q}})-1}
(z_3-z_4)^{({\vec{p}}{\vec{q}})-1}D^{ijk}C_{n_1n_2n_3}
({\vec{k}},{\vec{p}})\delta({\vec{k}}+{\vec{p}}+{\vec{q}})}}

Next, using the $SL(2,C)$ symmetry we can fix

$z_3=1,z_4=\infty,u=0$
where $u$ is again  the arbitrary point in the $c$-dependent correction term,
emerging in the BRST-invariant
expression of one of two integrated positively represented gluon operators 
in the
 (as we explained in the previous section).
Using (50) and (52) along with the momentum conservation
$\sum_{a=1}^4{\vec{p_a}}=0$
, the 4-point function is given by
\eqn\grav{\eqalign{
{A^{i_1...i_4}}({\vec{p_1}},
{\vec{p_2}},{\vec{p_3}},{\vec{p_4}})
={\prod_{j=1}^4}A^{m_j}({\vec{p_j}})\int_0^1dz_1\int_0^1dz_2\lbrace\cr
(z_1-z_2)^{(\vec{p_1}\vec{p_2})-1}(z_2-1)^{-1-({\vec{p_3}}{\vec{p_4}})}
C_{m_1m_2}^n
({\vec{p_1}},{\vec{p_2}})C_{m_3m_4n}({\vec{p_3}},{\vec{p_4}})\cr{\times}
(D^{i_1i_2}_jD^{ji_3i_4}+D^{i_1i_3}_jD^{i_2ji_4}+
D^{i_2i_4}_jD^{i_2i_3j})
+
(z_1-1)^{(\vec{p_1}\vec{p_3})-1}(z_2-1)^{-1-({\vec{p_2}}{\vec{p_4}})}
\cr{\times}C_{m_1m_3}^n
({\vec{p_1}},{\vec{p_3}})C_{m_2nm_4}({\vec{p_2}},{\vec{p_4}})
(D^{i_1i_2}_jD^{ji_3i_4}+D^{i_1i_3}_jD^{i_2ji_4}+
D^{i_2i_4}_jD^{i_2i_3j})\cr
+
(z_1-1)^{(\vec{p_1}\vec{p_4})-1}(z_2-1)^{-1-({\vec{p_2}}{\vec{p_3}})}
C_{m_1m_4}^n
({\vec{p_1}},{\vec{p_4}})C_{m_2nm_3}({\vec{p_2}},{\vec{p_3}})\cr\times
(D^{i_1i_2}_jD^{ji_3i_4}+D^{i_1i_3}_jD^{i_2ji_4}+
D^{i_2i_4}_jD^{i_2i_3j})\rbrace}}

Imposing the on-shell limit $({\vec{p_i}}{\vec{p_j}}){\rightarrow}0;
i,j=1,...,4$
it is straightforward to evaluate the integrals in $z_1$ and $z_2$,
so the final answer for the on-shell 4-point tree amplitude is

\eqn\grav{\eqalign{
{A^{i_1...i_4}}({\vec{p_1}},
{\vec{p_2}},{\vec{p_3}},{\vec{p_4}})={\prod_{j=1}^4}
A^{m_j}({\vec{p_j}})\lbrace
D^{i_1i_2}_jD^{ji_3i_4}+D^{i_1i_3}_jD^{i_2ji_4}+
D^{i_2i_4}_jD^{i_2i_3j}\rbrace
\cr\times\lbrace
{{C_{m_1m_2}^n
({\vec{p_1}},{\vec{p_2}})C_{m_3m_4n}({\vec{p_3}},{\vec{p_4}})}\over
{({\vec{p_1}}{\vec{p_2}})({\vec{p_3}}{\vec{p_4}})}}
+
{{C_{m_1m_3}^n
({\vec{p_1}},{\vec{p_3}})C_{m_2m_4n}({\vec{p_2}},{\vec{p_4}})}\over
{({\vec{p_1}}{\vec{p_3}})({\vec{p_2}}{\vec{p_4}})}}
\cr+
{{C_{m_1m_4}^n
({\vec{p_1}},{\vec{p_4}})C_{m_2m_3n}({\vec{p_2}},{\vec{p_3}})}\over
{({\vec{p_1}}{\vec{p_4}})({\vec{p_2}}{\vec{p_3}})}}\rbrace
\delta({\sum_{a=1}^4}{\vec{p_a}})
}}

This concludes the calculation of the 4-point amplitude of
the gluon-vertex operators. This amplitude is manifestly
cross-symmetric and the factors in numerators (quadratic in the
structure constants and hence quadratic in the momenta),
along with the transversality constraints on the polarization vectors,
protect it from the double poles. The group-theoretic factor
is easily recognized as $\sim{Tr(t^{i_1}...t^{i_4})}$
(with $t^{i_k}$ being the SU(3) generators), as one would expect 
for QCD amplitudes.
The bootstrap calculation performed in this section, based
on the zigzag invariance of the OPE of the gluons,
can be generalized to include higher number of points
and, in principle, the loop corrections as well.

\centerline{\bf Conclusions}
 
In this paper we have considered a new example of
gauge-string correspondence, constructing eight SU(3)
gluon vertex operators in RNS superstring theory
and showing them to reproduce perturbative 
QCD amplitudes.
Remarkably, the constructed 8 vertex operators
possess full zigzag symmetry and their OPE is closed. This leads to the absence
of intermediate massive states in their scattering amplitudes
which therefore have field-theoretic (rather than a stringy) structure,
reproducing the QCD dynamics.

It should be stressed here that the gauge-string isomorhpism
discussed in this paper is $not$ an analogue of AdS/CFT duality,
since the latter is the correspondence between strongly coupled
region of QCD and perturpative region of string theory in AdS
background. The example of gauge-string
correspondence constructed in this paper is the
one between perturbative QCD amplitudes and perturbative amplitudes 
in open string theory, similar to the correspondence
between D-instanton expansion in twistor string theory  
and perturbative expansion in $N=4$ super Yang-Mills theory,
observed by Witten ~{\wittwist} and elaborated 
in ~{\wb, \cachazo} and other works. 
It would be interesting to understand the connection between these
two examples of gauge-string isomorphism, as well as the relation
between twistor superstrings and RNS model.
Interestingly, twistor superstring theory with $SU(2,2|4)$
global symmetry emerges naturally in the context of 2T physics
{\barstwistt, \barstwist} and extra dimensions, where Berkovits-Witten theory
results as one of the holographic pictures of $4+2$-dimensional
string theory, obtained as a result of 2T gauge fixing.
We already have mentioned that the first order $\alpha$-
generators of $H_1\sim{H_{-3}}$ (related to the first hidden 
space-time dimension) are in one-to-one correspondence to the off-shell 
space-time symmetries observed in the 2T approach ~{\bars, \alfa}.
This appears to be an interesting project where many intriguing
connections can be anticipated.

The SU(3) colour group stems naturally from SU(3) subgroup
of the $\alpha$-generators, inducing global 
non-linear space-time isometries in hidden extra dimensions.
The gluon vertex operators are obtained as a result of
``photon painting'', i.e. by applying  SU(3) $\alpha$-generators to 
photon vertex operators. The zigzag invariance of gluon OPE is
ensured by the special property of 
the $\alpha$-symmetry, proven in this paper: 
the $\alpha$-transform of any massive vertex operator is trivial,
while the $\alpha$-generators applied to  photon produce new
physical states, corresponding to coloured QCD gluons.

This particularly suggests that each $\alpha$-generator of SU(3)
carries an associate colour-anticolour quantum number.
Since the $\alpha$-generators that paint a photon
can be classified in terms of
ghost cohomologies $H_{n}\sim{H_{-n-2}}$ while each cohomology 
is associated with hidden space-time dimension, 
this naturally implies that each hidden dimension has its own 
associate colour-anticolour contributing to the photon painting.
Attributing a colour-anticolour to a hidden dimension would nicely
match the classification of gluons in terms of ghost cohomologies.
Indeed, suppose the extra dimension associated with
$n=1$ carries  red and antired colours, the one 
related to $n=2$ paints with green and antigreen, the one of $n=3$ is 
blue-antiblue.  Let us start with $H_1\sim{H_{-3}}(n=1)$.
The only SU(3) $\alpha$-generator from this cohomology
is $L^{\alpha+}$, so we identify the corresponding
vertex operator $V^{\alpha+}$ with the $r{\bar{r}}$
(red-antired) gluon
The $n=2$ cohomology, associated with the second hidden dimension
has 3 generators, $L^{\alpha\beta}$ $L^{\beta+}$ and $L^{\beta-}$
The $\beta$-index adds the green (and anti-green) colour to the painting,
so the corresponding gluons are associated with the colour pairs
$g{\bar{g}},g{\bar{r}}$ and ${\bar{g}}r$.
Finally, the $n=3$ cohomology contains 4 vertex operators
$V^{\gamma\beta},V^{\gamma\alpha},V^{\gamma+}$ and $V^{\gamma-}$,
giving rise to 4 gluons with the colour combinations
involving the blue (and antiblue) colour:
$b{\bar{g}},{\bar{b}}g,b{\bar{r}}$ and ${\bar{b}}r$
(the $b{\bar{b}}$-pairing must be skipped since it is the linear
combination of the previously listed 
 gluons  $r{\bar{r}},g{\bar{g}}$
and the non-existing ``ninth'' gluon of white colour
$r{\bar{r}}+g{\bar{g}}+b{\bar{b}}$.

The construction discussed in this paper has so far involved the open
 string operators only. The role of the ghost cohomologies,
$\alpha$-symmetries
and their analogues  
in the closed string sector is yet to be understood.
Such an understanding would be 
particularly important since beyond the tree approximation
the Yang-Mills theory, with all the loops included, is described by
the closed string amplitudes.Clarifying the geometrical
meaning of closed string ghost cohomologies could particularly
connect our formalism and AdS/CFT approach.
Interestingly, the AdS/CFT correspondence generally does not fix a gauge group
which depends on various parameters on string theory side
(such as the number $N$ of RR-flux units in case of duality
between $SU(N)\times{SU(N)}$ gauge theory and type $IIB$ strings in warped
resolved conifold backgrounds ~{\klebgauge}. Our model, however, suggests
that the $SU(3)$ case is special, as it is related to the 
structure of extra dimensional $\alpha$-symmetries
classified the first three cohomologies. It is not clear at present if
the there are $\alpha$-symmetries in ghost cohomologies of  orders
higher than 3 and (if yes) if they have any geometrical meaning
like those of SU(3). 
If one proves that the set of $\alpha$-symmetries contained in the
first three cohomologies is complete, this may indicate that we have certain
new specific form of gauge-string duality in the case of $N=3$.
We hope to address these questions (along with many others)
in future works.

\listrefs
\end